\documentstyle[preprint,tighten,epsfig,twoside,floats,pra,aps]{revtex}
\begin{document}
\draft
\preprint{}
\title{
Two--Electron Effects in the Multiphoton Ionization of Magnesium 
with 400~nm 150~fs Pulses
}
\author{
D. Xenakis\(^{1,\mathrm{*}}\),
N. E. Karapanagioti\(^{1,}\)\thanks{Presently at: \textit{
Max--Planck--Institut f\"{u}r Quantenoptik, Hans--Kopfermannstrasse,
D-85748 Garching, Germany}}
and D. Charalambidis\(^{1,2}\)
}
\address{
\(^{1}\)Laser and Applications Division,
Institute of Electronic Structure and Laser,
Foundation for Research and Technology--Hellas,
P.O.Box 1527,
GR-711 10 Heraklion,
Greece
\\
\(^{2}\)Physics Department, University of
Crete, GR-713 10 Heraklion, Greece
}
\author{
H. Bachau
and E. Cormier
}
\address {
Centre Lasers Intenses et Applications (Universit\'{e} Bordeaux I-CNRS),
Universit\'{e} Bordeaux I, 
351 Cours de la Lib\'{e}ration, F-33405 Talence Cedex, France
}
\date{\today}
\maketitle
\begin{abstract}
The multiphoton ionization and photoelectron spectra of magnesium were
studied at laser intensities of up to \(6\times10^{13}\mathrm{~W
cm}^{-2}\) using 150~fs laser pulses of a wavelength of 400~nm. The
results indicated that a variety of different ionization mechanisms
played a role in both types of spectra. A theoretical model describing
the processes is presented and the routes to ionization are
identified. The work demonstrates the significance of the
two--electron nature of the atom in interpreting the experimental
results.
\end{abstract} 
%
%
\section{Introduction}
\label{sec:intro}

The behavior of atoms in intense electro-magnetic fields has been
under investigation since the advent of the laser.  Especially since
the development of short--pulse, high--intensity systems, experiments have
revealed a multitude of novel effects such as multiphoton ionization
(MPI)\cite{voronov1965}, above threshold ionization
(ATI)\cite{agostini1979}, harmonic
generation\cite{mcpherson1987,ferray1988} etc. Most of the theoretical
models originally proposed to describe the observations involved
essentially a single electron interacting with the electromagnetic
field. This approximation has led to significant understanding of the
atomic behavior in the field.  However, the multi-electron nature of
most atoms can play a significant role under certain circumstances.
Theory then has to take into account multi--electron excitation,
correlations between the electrons, multiple continua, i.e. the
detailed atomic structure in order to be able to predict the atomic
behavior.  This fact emerges as both theoretical and experimental
techniques become more advanced, enabling the exploration of new
regimes and the detailed testing of theoretical models.

In multi-electron atoms, significant double excitation can take place
leaving the ion not in the ionic ground state but in an excited state
or possibly ejecting a second electron.  In the alkaline earth atoms
for example, the existence of doubly excited-state manifolds between
the two ionic thresholds alters drastically the single-electron
picture rendering it unsuitable. The conditions under which these
effects may occur are easily within reach of present experimental
setups. Past work by Dalwoo Kim et al \cite{dalwookim1990}, DiMauro et
al \cite{dimauro1988} and van Druten et al \cite{vandruten1994} has
looked into the multiphoton ionization of magnesium with nanosecond
and picosecond laser pulse durations and indicated the importance of
short pulse experiments in understanding the processes involved. At
the same time, theoretical methods incorporating these effects are
also being developed \cite{tang1990,zhang1996}, creating the need for
further experimental investigations.

In the present work we seek to investigate the extent to which
electron correlations play a role in the high intensity regime for
multi-electron atoms.  We chose magnesium, a fairly low atomic weight
alkaline earth element, as it is accessible to both experimental and
theoretical investigation.  We report measurements of the multiphoton
single and double ionization and ATI of magnesium using 150~fs, 400~nm
laser pulses.  A theoretical model has been developed and applied
which describes the processes leading to ionization and shows that the
inclusion of one and two-electron effects are necessary for
an accurate description.  Measurements of the multiphoton ionization
yield displayed the expected perturbative behavior for the production
of the Mg$^+$ ion but showed unexpected features in the double
ionization yield.  The ATI electron spectra also exhibited complex
structure, originating from a variety of single and double ionization
mechanisms, whose relative importance changes as the intensity is
varied.  The theoretical treatment involved both a perturbation theory
description including all correlation effects between the active
electrons and additionally, for the ionization of Mg$^+$, a
non-perturbative approach.

\section{Method}
\label{sec:method}

\subsection{Experimental set-up}
\label{sec:experimentalsetup}

\begin{figure}
\centerline{\epsfig{figure=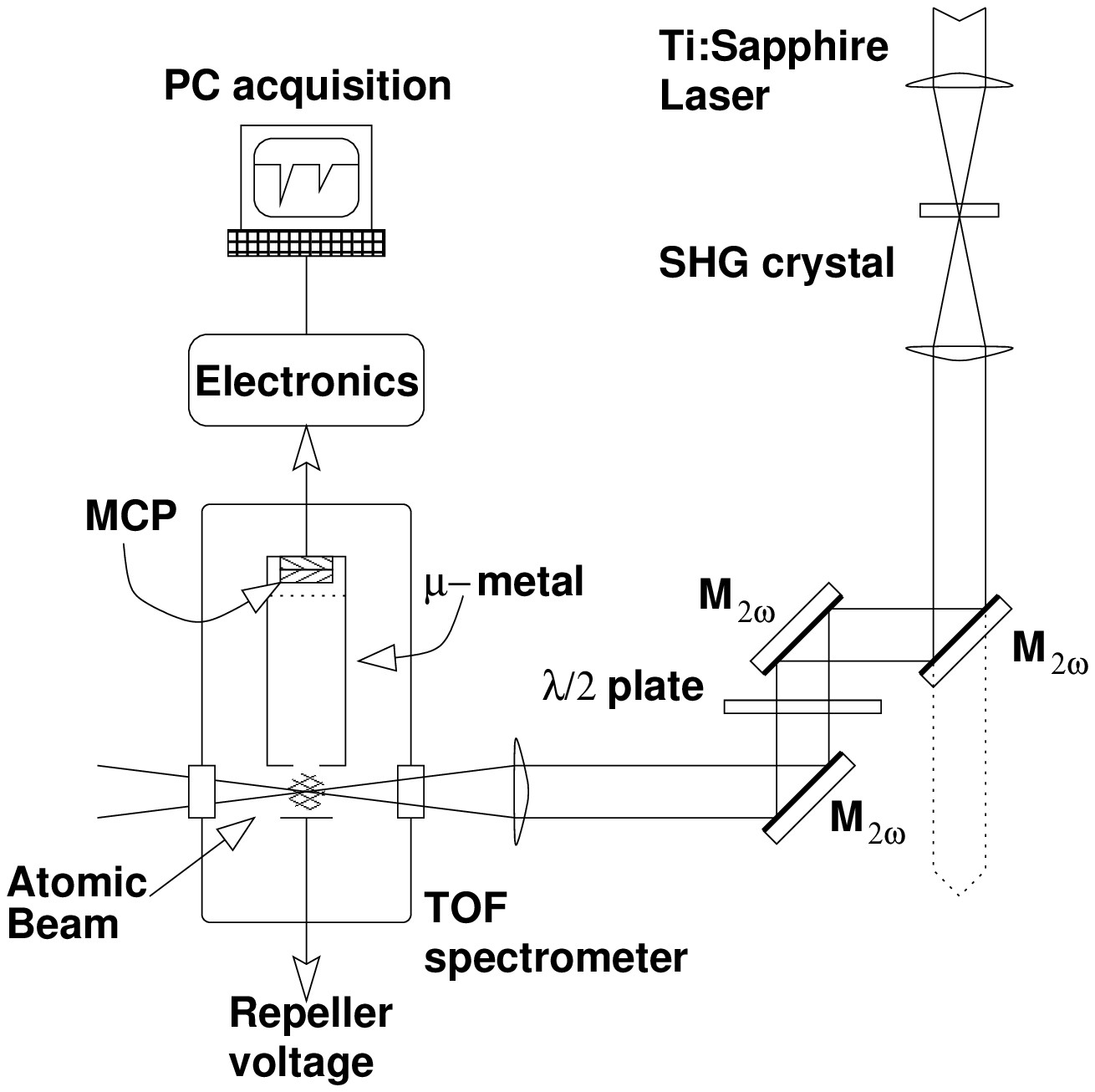,width=0.45\textwidth}}
\caption{ Experimental setup. SHG: Second harmonic generation,
$\mathrm{M}_{2\omega}$: mirror for the second harmonic radiation, TOF:
Time-of-flight, MCP: Micro-channel plates.  }
\label{fig:exptlsetup}  
\end{figure}

A plan of the experimental set-up is shown in figure~\ref{fig:exptlsetup}.  
The laser employed consists of a titanium
sapphire oscillator amplified in a regenerative amplifier utilizing a
chirped pulse amplification technique \cite{strickland1985}. The laser
operates at 1 kHz repetition rate, with a measured pulse duration of
$\approx$150~fs at a wavelength of $\approx$798~nm. The second
harmonic of this beam was then produced using a focused geometry in a
250~$\mu$m thick BBO crystal. The harmonic radiation was separated
from the co-propagating fundamental by employing three dichroic
mirrors that were transparent to the fundamental (denoted by
\(\mathrm{M_{2\omega}}\) in figure~\ref{fig:exptlsetup}).  The
harmonic generation efficiency of this setup was $\approx$10\%. The
laser energy of the fundamental was set to be less than 300~$\mu$J,
thus producing a maximum of $\approx$30~$\mu$J of second harmonic
radiation.

The resulting harmonic laser beam was focused in a time--of--flight 
spectrometer (TOF) by a MgF lens of 15~cm focal length.  A
half--wave plate was positioned prior to that in order to
align the polarization of the light with the TOF tube axis. Under
the conditions of the experiment, the lens did not affect
significantly the beam either in terms of non-linear absorption or by
pulse-front curvature at the focus \cite{bor1989}. The energy
measurement was achieved with a fast vacuum photodiode which was
cross-calibrated with an average power meter.

The time--of--flight spectrometer, which has a field--free flight tube
of 15~cm oriented perpendicular to the laser propagation axis, is
equipped with a $\mu$-metal magnetic shield. This shield, in
combination with a repeller plate and the appropriate polarity of the
charged particle detector, allowed the use of the instrument
alternately as an electron spectrometer and as an ion time--of--flight
system.  For the electron energy measurements the repeller was
grounded, whereas for the operation of the TOF as a mass spectrometer
the repeller was charged with a voltage of up to +1000~V. The charged
particle detector at the end of the flight tube was a double
micro--channel plate in a chevron configuration. Its front side was
grounded for detecting electrons and at a negative voltage for
ions.

An oven was inserted at the top of the interaction chamber filled
with magnesium pellets which, when heated, produced an atomic beam as
the vapor exited an aperture. The atomic beam met the laser beam and
the time--of--flight tube at right angles.

The acquisition system was essentially the same for both electron and
ion detection. The signal from the micro--channel plates first went through
a fast amplifier and a discriminator. 
Subsequently, its time of arrival with respect to the laser pulse was
recorded by a multiple--event time--digitizer card which was installed
in a personal computer. The main feature of this card was that it could
acquire and time one event per 0.5~ns time--bin with no dead--time
between bins. Combined with the bandwidth of the cabling and
electronic processing stages, the effective time resolution of the
system was approximately one nanosecond. This acquisition system was
not sensitive to the amplitude of the signal and hence it was operated in
the event--counting regime. This meant that
the number of events \textit{in each bin} had to be less than one per
laser pulse, a condition which was ensured throughout the measurement.

At the higher laser intensities, for which the counting method was no
longer appropriate due to the increased ionization rates, the analogue
signal was obtained directly from the micro-channel plates and
recorded on a digital oscilloscope. For the ion yield versus laser
intensity spectra, where data were recorded using both methods of
acquisition, the data were matched at several points in an
intermediate regime. They were further cross-checked with data from
the less abundant isotopes of magnesium which would be still in the
counting regime at intensities for which the main peak was
not. Lastly, to ensure that no space--charge effects were distorting
the data, measurements were taken at different oven temperatures (and
consequently atomic number densities) and compared.  No rigorous
absolute energy calibration of the electron spectrometer was
performed. The calibration of the energy scale was based on the
expected inter-peak separation of the several ATI peaks.  This
resulted in a determination of the relative peak positions to better
than 0.1~eV, although the determination of the absolute peak positions
was hindered by uncertainty in the zero time.  Comparison to data with
xenon was consistent with the magnesium results in so far as the
energy calibration of the spectra was concerned, confirming the
validity of the calibration. Xenon was also used to estimate the
intensity of the laser at the interaction region
\cite{chara1997,uiterwaal1998,perry1988} and compare it with the
number obtained by estimating the focal spot size.

\subsection{Theoretical modeling}
\label{sec:theory}

\begin{figure}
\centerline{\epsfig{figure=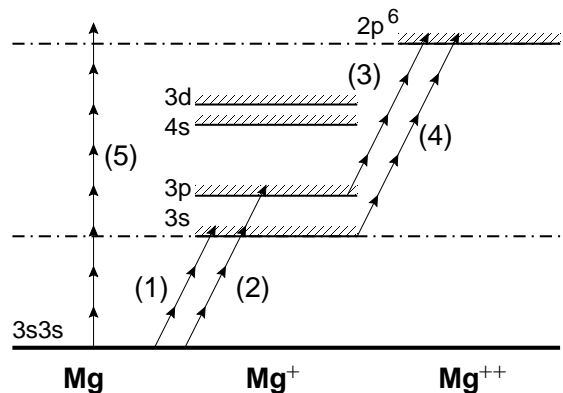,width=0.45\textwidth}}
\caption{
Partial energy level diagram of magnesium and the relevant transitions 
involving 400~nm photons in the neutral atom (Mg) and its ionic states
(Mg$^+$ and Mg$^{++}$). The different paths to ionization indicated by 
numbers (1) to (5) are described in the main text and in
Table~\ref{tbl:mglevels}.
}
\label{fig:mgleveldiagram}   
\end{figure}

We now propose a theory to describe the experimental results and
explain most of the essential features. Magnesium in the present
context is a rather complicated system since it involves many coupled
states and different ionization channels. However, in that case, we
were able to identify and isolate atomic states and transitions
connected to the leading processes. These elements are summarized in
figure~\ref{fig:mgleveldiagram}. We consider that the electrons
measured in the experiment are produced by 4 different processes. Two
lead to the production of Mg$^{+}$ (single ionization to the ground
state Mg$^{+}$(3s) by absorption of 3 photons or to the first exited
state Mg$^{+}$(3p) by absorption of 4 photons) and are labeled process
(1) and (2) respectively. Two other processes sequentially double
ionize the system from the ion Mg$^{+}$. Mg$^{++}$ is thus produced by
absorption of 4 photons from Mg$^{+}$(3p) or by absorption of 5
photons from Mg$^{+}$(3s).  The basic assumption in the theoretical
approach is that double ionization is produced through a sequential
process. Therefore, double ionization through direct (coherent)
absorption of 8 photons (see path (5) in figure~\ref{fig:mgleveldiagram}) 
is neglected. Under these conditions, the
population of each species is obtained by solving the following rate
equations:

\begin{eqnarray}
\label{eq:rate_equation}
\dot{N}_{3s^2} & = &-(\Gamma_{3s^2\rightarrow 3sk_1}^{(3)}+
                      \Gamma_{3s^2\rightarrow 3pk_2}^{(4)})\ 
                      N_{3s^2}  \nonumber \\ 
\dot{N}_{3s}   & = &\Gamma_{3s^2\rightarrow\ 3sk_1}^{(3)}\ N_{3s^2}-
                    \Gamma_{3sk_1\rightarrow k_4k_1}^{(5)}\ 
                    N_{3s} \\ 
\dot{N}_{3p}   & = &\Gamma_{3s^2\rightarrow\ 3pk_2}^{(4)}\ N_{3s^2}-
                    \Gamma_{3pk_2\rightarrow k_3k_2}^{(4)}\ 
                    N_{3p} \nonumber
\end{eqnarray}
with the initial conditions ${N}_{3s^2}(t = -\infty)=1$ and all other
populations set to zero. ${N}_{3s^2}$, ${N}_{3s}$ and ${N}_{3p}$
represent the time-dependent populations of Mg(3s$^2$), Mg$^{+}$(3s)
and Mg$^{+}$(3p) respectively.  The time-dependent width
$\Gamma_{3ln_1\rightarrow n_2n_3}^{(p)}$ refers to the $p$-photon
transition from the initial state $3ln_1$ to the final state $n_2n_3$.
When the specification of the angular momentum has been removed from
the state notation (like in $n_2n_3$), the width includes the
contribution from each accessible symmetry.  $k_n$ refers to the
momentum of the electron ejected through the process labeled ($n$) in
figure~\ref{fig:mgleveldiagram}. The electron energies 
(reported in table~\ref{tbl:mglevels}) are deduced from
the energy conservation principle excluding laser induced level
shifts. The atomic structure calculations providing the atomic energy
levels and the transition cross-sections have
been described in detail in a precedent paper
\cite{karapanagioti1996b}.  The energies and wave-functions are
computed using a frozen-core configuration interaction (CI) procedure.
The core Mg$^{++}(1\mathrm{s}^{2}2\mathrm{s}^{2}2\mathrm{p}^{6})$
(where electrons are assumed ``frozen'') is represented by a
self-consistent-field wave function. The core potential (including a
dielectronic polarization term) is first used to determine Mg$^{+}$
orbitals. We then compute two-electron wave-functions by diagonalizing
the 2-electron Hamiltonian in a basis of Mg$^{+}$ orbitals
configurations. The multiphoton ionization rates
$\Gamma_{3s^2\rightarrow 3lk}^{(p)}$ and $\Gamma_{3lk \rightarrow
  k'k}^{(p)}$ are calculated within lowest order perturbation theory
\cite{bachau1998} (for the calculation of MPI matrix elements see also
\cite{cormier1995} and references therein). The rates leading to
double ionization from Mg$^{+}$ are deduced from non-perturbative ATI
calculations \cite{cormier1997} (see below) and are in agreement with
the perturbative results. The spatio-temporal dependence of the laser
is accounted in this model by assuming the following field envelope:

\begin{equation}
\label{eq:spatiotemporaldep}
I(r,t)=I_{max}\ \frac{1}{(cosh(1.76\frac{t}{\tau}))^2}\ \
       e^{-(4\ln 2(\frac{r}{\kappa})^2)}
\end{equation}
where we assume a cylindrical symmetry $(r,z)$ for the space macro
dependence of the intensity. $\kappa$ and $\tau$ are respectively the
spatial and temporal full width at half maximum (FWHM).

Besides the perturbative evaluations of ionization rates, we have
performed non-perturbative calculations of ionization from
Mg$^{+}$(3s) and Mg$^{+}$(3p) by numerically solving the time
dependent Schr\"odinger equation (TDSE)\cite{cormier1997}. A
pseudo-potential \cite{schwerdtfeger1982} is used to model the Mg$^+$
ionic potential. The total wave-function is expanded in terms of
spherical harmonics and radial B-spline functions. The total
Hamiltonian is then propagated in time and the electron spectrum is
calculated at the end of the pulse. This approach has the advantage of
neglecting none of the couplings. It therefore accounts for all
dynamical level shifts and possibly resonant transitions. These latter
effects are of importance to interpret some of the experimental
results as it is discussed below.

\section{Experimental results and comparison with theory}
\label{sec:exptlresults}

\subsection{Ion Yields}
\label{sec:exptresultsion}

\begin{figure}
\centerline{\epsfig{figure=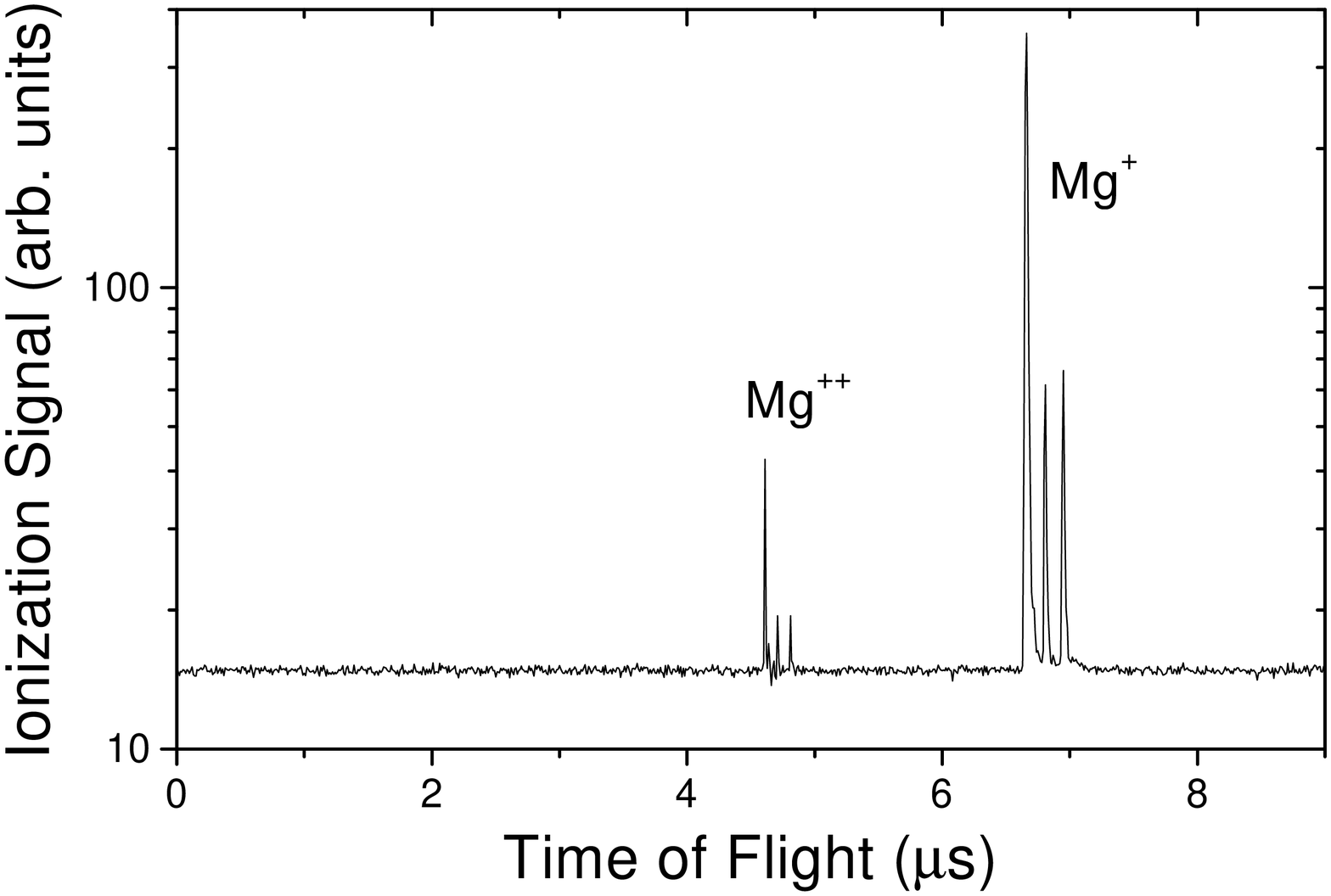,width=0.45\textwidth}}
\caption{
Typical time of flight spectrum obtained during the experiment. While
the singly and doubly ionized magnesium species can be observed, no
impurities are evident. The isotopic ratios are as expected (A=24:
78\%, 25: 10\%, 26: 11\%). Note the logarithmic scale.
}
\label{fig:iontofspectrum}     
\end{figure}

Figure~\ref{fig:iontofspectrum} shows a typical time--of--flight ion
mass spectrum obtained in the experiment, in logarithmic scale.  The
two charge states of magnesium (Mg$^{+}$ and Mg$^{++}$) are evident,
each displaying peaks corresponding to its three dominant
isotopes.  The isotope ratios are as expected and no presence of
impurities is detected. By integrating under the main peaks of such
spectra obtained for different laser energies, the ion yield curves of
singly and doubly ionized magnesium were obtained.
 
\begin{figure}
\centerline{\epsfig{figure=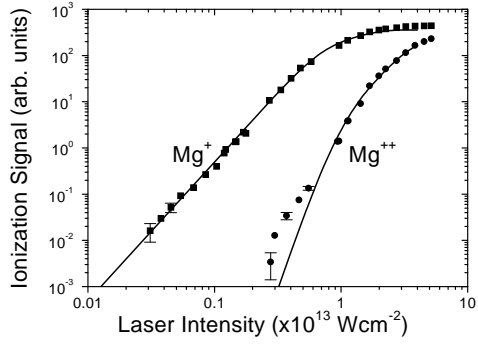,width=0.45\textwidth}}
\caption{
Experimental and theoretical ion yields. 
The square data points correspond to Mg$^+$
whereas the circles to Mg$^{++}$. The solid lines are the calculated
yields (rate equations and perturbation theory). 
}
\label{fig:ionyields} 
\end{figure}

The resulting ion yield curves are shown in figure~\ref{fig:ionyields}. 
The theoretical curves (given by solving
equation~\ref{eq:rate_equation} and denoted by the solid lines) agree very well
with experimental values over most of the range investigated. Small
discrepancies in the heavily saturated region above
\(2\times10^{13}\mathrm{\ W~cm}^{-2}\) are expected since the detection
volume was different to the interaction volume. The precise experimental
intensity was determined by the agreement with the theoretical
curves. However, some uncertainty to this estimate can  be expected due to the
fact that the macro-dependence of the field envelope assumed in the
theory strongly affects the ion yield behavior, especially in the saturated
region, sometimes even affecting the apparent saturation intensity.
The theoretical rate for three photon ionization of Mg is
$\Gamma_{3s^2\rightarrow 3sk_1}^{(3)} = 0.45\times 10^{6}\ 
(\frac{I}{I_0})^3$ a.u. (where $I_0=6.44\times 10^{15}\mathrm{\ 
  W~cm}^{-2}$). The corresponding generalized cross-section
$\sigma_{3s^2\rightarrow 3sk_1}^{(3)} = 0.85\times 10^{-80}
\mathrm{~cm^{6}~sec^{2}}\ $ is in agreement with the values calculated
by Chang and Tang \cite{chang1992}.  One can easily calculate that
single ionization saturates at about $I=5\times 10^{12}
\mathrm{~W~cm^{2}}\ $ for a pulse duration of 150~fs, in agreement with
the experimental ion yield shown in figure~\ref{fig:ionyields}.

In the lower energy range of the figure (around $I=4\times
10^{12}\mathrm{\ W~cm}^{-2}$) a clear enhancement to the Mg$^{++}$
yield relative to what expected from the theoretical results can be
seen. This enhancement seems to level off, the experimental curve
tending back to the theoretical one, at the onset of saturation of
single ionization.  This may be an indication that two--electron
ejection is connected to the population of the atomic ground state
which would be the case for e.g.  non-sequential double ionization.
It should be noted that the Keldysh or adiabaticity
parameter\cite{Keldysh1965} \(\gamma = \sqrt{I_{p} / (2 U_{p})}\)
(where $I_{p}$ is the ionization potential of the atom and \(U_{p}
\approx 9.33 \times 10^{-14} I(\mathrm{W~cm}^{-2}) \lambda^2
(\mathrm{\mu m}) \) is the ponderomotive energy of a free electron in
a laser field of wavelength $\lambda$ and intensity $I$) is not
favorable to observe tunneling, rescattering or shake-off effects
($\gamma = 2$ at the highest intensity). It is therefore possible that
the enhancement is related to atomic structure, or to dynamical shifts
leading to either resonances or channel closing as it is discussed in
section \ref{sec:electronshifts}.

\subsection{Electron results}
\label{sec:exptresultslelectron}

\begin{figure}
\centerline{\epsfig{figure=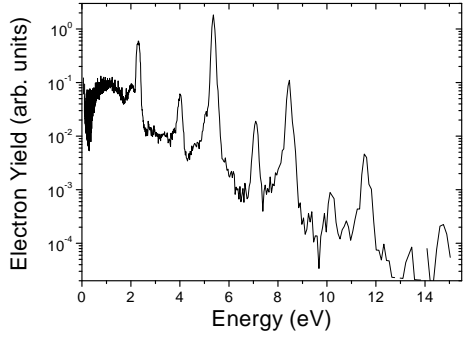,width=0.45\textwidth}}
\caption{
Electron spectrum at a laser intensity of 
$2\times 10^{13} \mathrm{Wcm}^{-2}$. Two dominant sets of electron
peaks are observable.
}
\label{fig:electronspectrum} 
\end{figure}

A typical electron energy spectrum is shown in 
figure~\ref{fig:electronspectrum}.  The spectrum displays a set of main ATI
peaks, each accompanied by a pronounced side peak (at approximately
1.3~eV to its low energy side). While the main ATI peak can
safely be attributed to three-photon ionization of neutral magnesium
leading to the ion Mg$^{+}$(3s), the origin of the side peak is not so
clear without further investigation.  Note that the main peaks in
figure~\ref{fig:electronspectrum} are offset compared to the expected
energies corresponding to normal perturbation theory.  This is an
experimental artifact due to the lack of absolute energy calibration
of the TOF as it is not relevant to the present investigation (see
also section~\ref{sec:experimentalsetup}).

Table~\ref{tbl:mglevels} gives the calculated energy of the ejected
electrons upon ionization through the different processes depicted in
figure~\ref{fig:mgleveldiagram}.  Comparing these with the electron
spectra of figure~\ref{fig:electronspectrum} suggests two possible
mechanisms that would result in the side peak in the recorded data:
(i) 4--photon ionization of Mg leading to the Mg$^{+}$(3p) threshold,
i.e. the first excited state of the magnesium ion (process (2) in
table~\ref{tbl:mglevels}).  That would give an energy difference
between the main and side peak of $(1.656 - 0.34)=1.316$\ eV, (ii)
5--photon ionization of Mg$^{+}$(3s) leading to the 2p$^6$ ground state of
Mg$^{++}$ ((4) in table~\ref{tbl:mglevels}), which would give an energy
difference of $(1.656 - 0.47)=1.186$\ eV. Both of these candidates are
within the experimental uncertainty and the electron peak widths,
especially at the higher intensities.  Note that direct double
ionization (process (5) in table~\ref{tbl:mglevels}) would give a
continuous electron distribution with maximum energy of 2.13~eV, as
that energy is shared between the two electrons.

\subsubsection{Electron shifts}
\label{sec:electronshifts}

\begin{figure}
\centerline{\epsfig{figure=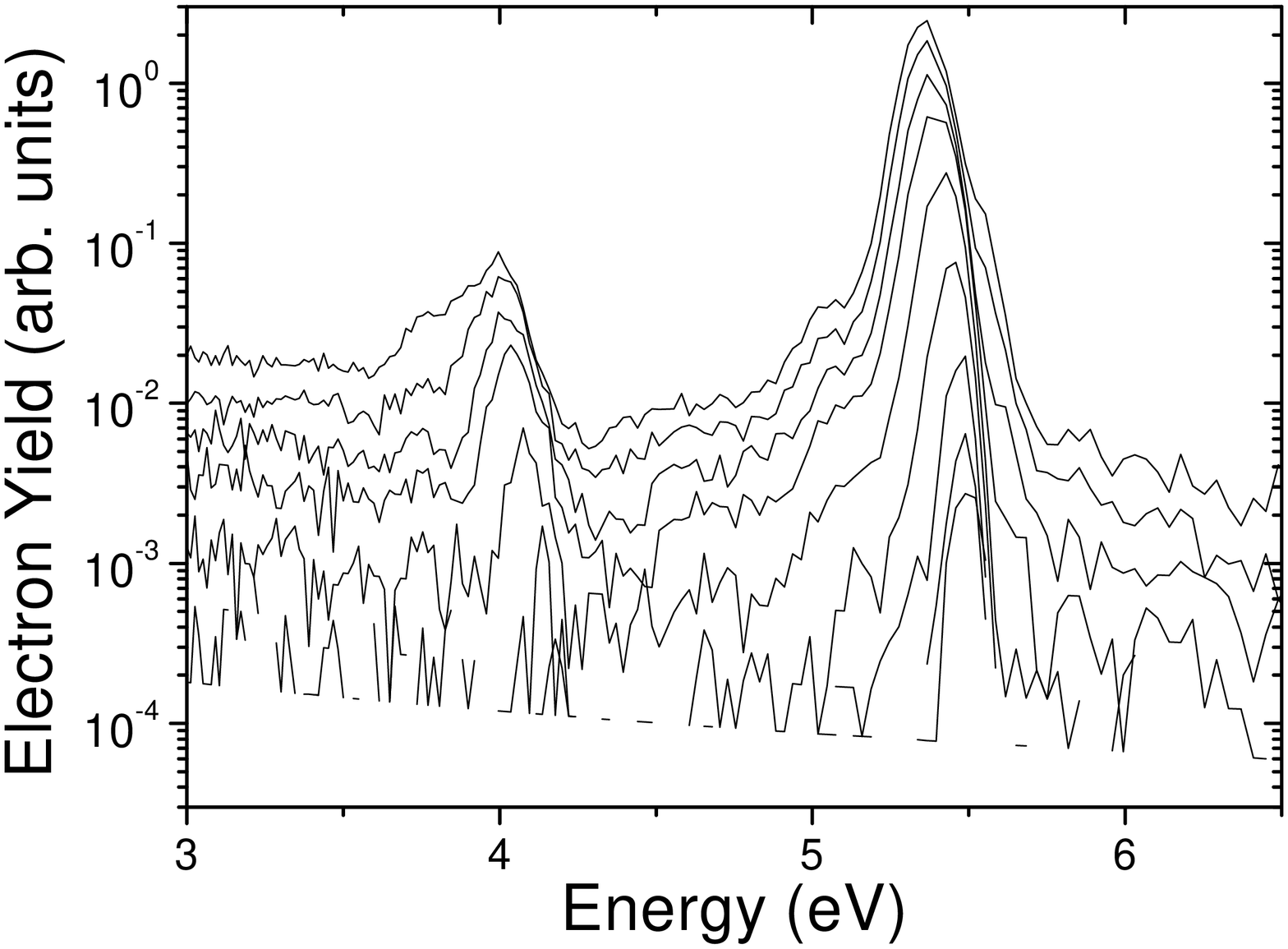,width=0.45\textwidth}}
\caption{
Enlarged section of the electron energy spectra of figure 
\ref{fig:electronspectrum} focusing on a
main peak and a sub-peak, for various laser intensities. A shift of
the peaks towards the higher energies as the laser power decreases is
observable. (Intensities from top to bottom in $10^{13}$~W~cm$^{-2}$: 
4.7, 3.6, 2.3, 1.9, 1.4, 0.7, 0.4, 0.24, 0.22).
}
\label{fig:electronspectrumdetail} 
\end{figure}

\begin{figure}
\centerline{\epsfig{figure=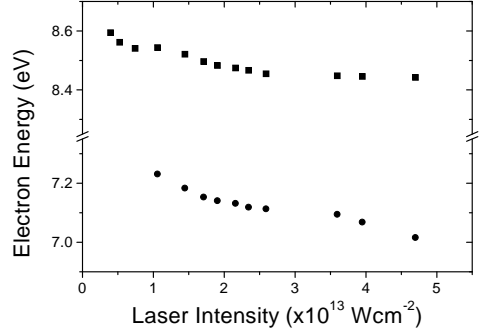,width=0.45\textwidth}}
\caption{
Electron peak positions vs. laser intensity for a pair of peaks at 
approximately 8.5~eV and 7.1~eV. Note the axis break.
(see text for more details).
}
\label{fig:elecpeakvsint}
\end{figure}

Careful analysis of the electron spectra can give more insight into
the mechanisms involved. Figure~\ref{fig:electronspectrumdetail} shows
a detail of the ATI spectra of figure \ref{fig:electronspectrum} with
additional plots for different intensities. If one concentrates just
on the peak positions one notices that the main peak shifts smoothly
up to a point and then stops completely. In contrast, the side peak
initially shifts, then stops and finally starts shifting again. This
behavior is shown clearly in figure~\ref{fig:elecpeakvsint} where the
positions of two electron peaks are plotted versus laser
intensity. The square points correspond to the main peak while the
circles to the side peak.

The main peak originates from the processes (1) and (3) shown in
figure~\ref{fig:mgleveldiagram}. Our calculations show that, in all
cases, the contribution from process (1) (three-photon ionization
leading to a Mg$^{+}$(3s) ion) dominates over (3).  Indeed, this
latter process originates from Mg$^{+}$(3p), whose production from
Mg(3s$^2$) requires the absorption of 4 photons in contrast to the 3
photons involved in process (1).  The flattening can be explained in
terms of the ion production saturating
\cite{chara1997,uiterwaal1998}. As the intensity increases, there
comes a point where, in effect, the atom will ionize before the end of
the laser pulse, and so it will not experience the full power of the
laser.  Therefore the curve levels out.  To account for the specific,
detailed behavior one has to take into account all the factors that
would contribute to the shift, e.g. (i) ponderomotive shifts and (ii)
AC Stark shifts.

\begin{figure}
\centerline{\epsfig{figure=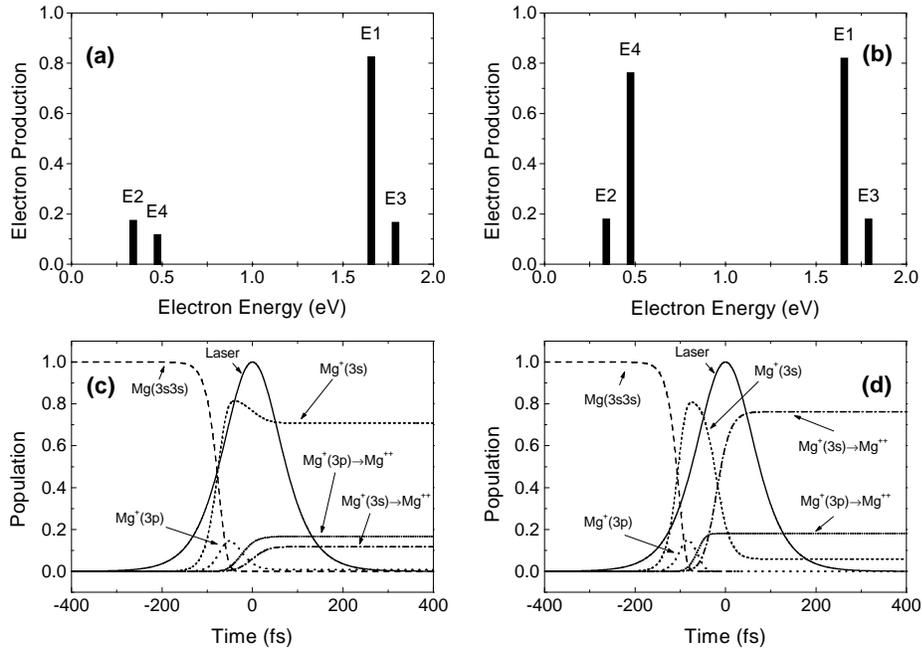,width=0.9\textwidth}}
\caption{Ion and electron populations calculated with no intensity
space macro-dependence at two different intensities.  (a), (b)
Electron production at intensities of 
$I=1.7\times 10^{13}\ \mathrm{W~cm}^{-2}$ and 
$I=3\times 10^{13}\ \mathrm{W~cm}^{-2}$
respectively. Electron peaks correspond to processes 1-4, as indicated
in figure \ref{fig:mgleveldiagram};  (c), (d) Time dependent
populations of different ionic species at the intensities
corresponding to cases (a) and (b) as the laser pulse evolves.
(solid line): Laser intensity normalized to the maximum intensity; 
(dashed line): ground state of the atom Mg(3s$^2$)
(short-dash line): ground state of the ion Mg$^+$(3s) produced by 
three-photon ionization
from Mg(3s$^2$) along with electron of energy E1;
(dotted line): first excited state of the ion Mg$^+$(3p) produced by 
four-photon ionization
from Mg(3s$^2$) along with electron of energy E2;
(dash-dot-dot line): doubly ionized Mg$^{++}$(2p$^6$)
through four-photon ionization of Mg$^+$(3p) along with electron of
energy E3;
(dash-dot line): doubly ionized Mg$^{++}$(2p$^6$)
through five-photon ionization of Mg$^+$(3s) along with electron of
energy E4;
}
\label{fig:elecpopulations}
\end{figure}

\begin{figure}
\centerline{\epsfig{figure=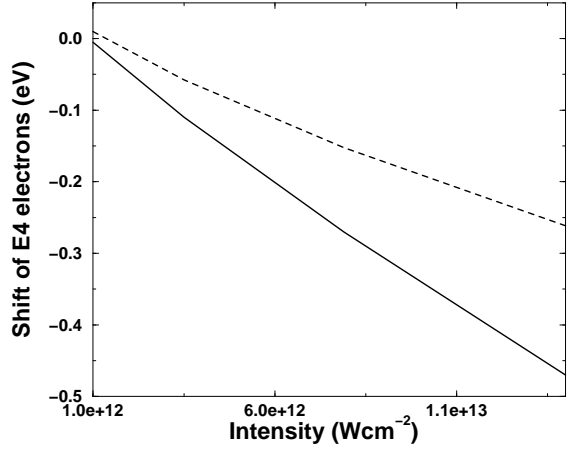,width=0.45\textwidth}}
\caption{Total shift (solid line) in energy of electrons originating 
from process (4), resulting from a non-perturbative calculation of the 
ionization of Mg$^+$(3s). Also shown (dashed line) is the contribution
to the total due to the Mg$^+$(3s) AC stark shift.}
\label{fig:elecpeakvsintth}
\end{figure}

The lower data points in figure~\ref{fig:elecpeakvsint} (corresponding
to the side peak) have a different behavior, which cannot be explained
solely by saturation. The two different trends observable in the
energy shift indicate that more than one process is
at work.  As can be seen from the theoretical calculations in
figure~\ref{fig:elecpopulations}(a) and (b), the side peak is a
(unresolved in our data) combination of two processes, one coming from
single ionization (2) the other from double sequential ionization (4)
(see figure \ref{fig:mgleveldiagram}).  At low intensity
(figure~\ref{fig:elecpopulations}(a)), the calculations show that the
side peak is dominated by single ionization (2).  At about
$I=1.8\times 10^{13}\ \mathrm{W~cm}^{-2}$, processes (2) and (4) have
an equal contribution, but for higher intensities the process (4)
largely dominates over (2) (figure~\ref{fig:elecpopulations}(b)).
Figure~\ref{fig:elecpopulations}(c) and (d) indicates the development
of the relevant atomic populations as the laser pulse evolves.  In
fact, as we will see below, the process (2) should have a
significantly smaller contribution for intensities larger than
$I=8\times 10^{12}\ \mathrm{W~cm}^{-2}$.  The electron production
originating from the decay to the first excited ionic state would
saturate at the same intensity as the electrons of the main peak, in
agreement with the first trend observed in the data. On the other
hand, the electrons originating from sequential double ionization (4)
are expected to saturate at higher intensities.  As the intensity
increases, this second process dominates in the side peak, and the
change of slope observed after the initial leveling off is due to the
shift experienced by this second different peak. This latter shift is
shown in figure~\ref{fig:elecpeakvsintth} where we plot the energy
shift of electrons originating from process (4) versus the
intensity. This graph results from a non-perturbative calculation
which includes both AC Stark shift and ponderomotive shift. Note that
the shift is not uniquely due to the ponderomotive effect as is
often the case in ATI measurements. This is shown in
figure~\ref{fig:elecpeakvsintth} where we detail the contribution of
the ponderomotive shift to the total shift. It should be noted that
the experimental shifts appear to have a smaller magnitude than the
theoretical ones, indicating a smaller experimental intensity than
previously extracted from the ion yields. However, considering that
the acquisition of the electron signal takes place under different
conditions to that of the ion signal (field-free, as opposed to a high
applied voltage for the ion extraction), it is extremely likely that
the extraction region, and therefore the effective intensity seen by
each species could well be different. The uncertainty in the
determination of the intensity in the ion yields should also be taken
into account, as described in a previous section.

Our calculations show that, for intensities larger than $10^{13}\
\mathrm{W~cm}^{-2}$, the double ionization is dominated by the process
(1+4) (see figure~\ref{fig:mgleveldiagram}). At lower intensities, the
channel (2+3) has a significant contribution to the production of
Mg$^{++}$.  As already mentioned, the double ionization yield displays
a significant deviation at low intensities. We considered the laser
induced energy shift of the Rydberg series Mg$(3pnl)$ (converging, as
$n \rightarrow \infty$, to the sum of the AC Stark shift of
Mg$^{+}$(3p) and the ponderomotive shift $U_p$) and the AC Stark shift
of Mg(3s$^2$). Through this we found that, for intensities larger than
$8\times 10^{12}\ \mathrm{W~cm}^{-2}$, the 4-photon ionization channel
(2) is closed and ionization to Mg$^{+}$(3p) now requires the
absorption of 5 photons.  The effect is that the experimental double
ionization yield deviates from the normal picture at about $I=8\times
10^{12}\ \mathrm{W~cm}^{-2}$. Note that the single ionization channel,
dominated by the process (1), is almost not affected by this effect.
Nevertheless, we see in figure~\ref{fig:ionyields} that the
theoretical Mg$^{++}$ yield lies below the experimental curve in the
low intensity region.  According to the previous considerations, it
should fit the Mg$^{++}$ experimental yield at low intensity and
overestimate it for $I>8\times 10^{12}\ \mathrm{W~cm}^{-2}$ (in the
rate equation system~(\ref{eq:rate_equation}), 4-photon ionization of
Mg to Mg$^{+}$(3p) is allowed at all intensities). Thus the
discrepancy between theoretical and experimental results in the latter
region still remains to be explained.  Other effects, like direct
and/or sequential ionization (eventually involving higher thresholds)
enhanced by the presence of resonances, should be
explored. Consequently, the sequential model adopted here for double
ionization may not be valid at low intensities, where further
experimental and theoretical efforts may be required in order to
complete the present analysis.

\section{Conclusions}
The results presented in this paper are some of the very few works
studying in both experimental and theoretical detail the MPI and ATI
spectra of an alkaline earth using ultra--short pulses. The work
demonstrates clearly the significance of the two--electron nature of
the atom in these spectra, both in terms of the role of the electron
correlations, and with respect to the precise level structure of both
the atom and the ion.

The attribution of the pronounced sets of side peaks, accompanying the
usual ATI spectrum, was not trivial, and required a theoretical
investigation taking into account all of the above factors in order to
explain their origin. As it turned out, more than one processes
contributed to their appearance, each dominating at a different range
of intensities. Most notably, one of the two dominant processes at low
intensity was the decay to the first excited electronic state of the
ion, a pathway which involves the excitation of both valence electrons
and coupled ionization channels.  However, the prominence of this
process was diminished as the intensity increased due to the depletion
of the ground atomic state, and overtaken by the electron originating
from sequential double ionization from the ground state of the atom.
In principle, another new peak should arise due to sequential double
ionization from the excited ionic state, but the position of this was
calculated to be too close to the main structure (three-photon
ionization of Mg) to allow for its observation. It might be useful
to conduct an extension of this study to a different photon energy, 
at the moment not available in our laboratory, at
which these proximities are avoided, and at which the peaks from each
of the processes are well resolved.

The detailed studying of the electron energy shifts of both the main
and the side set of peaks also pointed to the importance of the
factors mentioned above. Quite apart from their being used as a tool
towards the identification of the side peak, the shifts displayed a
behavior not sufficiently explained by the common analysis in terms of
ponderomotive shifts. Due to the complex level structure of both the
atom and the ion, with transitions close to resonance with the laser
field, AC Stark shifts were found to play an important role and could
no longer be ignored.  The multiphoton ionization yields, despite the
relatively low intensities employed, also display an interesting
enhancement of double ionization which deviates from the normal
picture.  We have found that, due to AC and ponderomotive shifts, the
channel (2) requires the absorption of 5 photons for intensities
larger than $8\times 10^{12}\mathrm{W~cm}^{-2}$.  This is a possible
explanation for the enhancement of double photon ionization, but other
effects should be explored, in particular the possibility of
near-resonant enhancement or, more tentatively, direct double
ionization, possibly enhanced by electron correlations and presence of
multiple doubly excited manifolds.

All of the above indicate that a full picture of the atom needs to be
explicitly contained in any investigation attempting to elucidate the
high-intensity behavior of two electron atoms, and possibly other
atoms as well.

\acknowledgments
\label{sec:acknowledgments}

This work has been carried out in the Ultraviolet Laser Facility of
FORTH with partial support from the TMR Program of the European Union
(contract number ERBFMGECT950017). We gratefully acknowledge useful
discussions with Prof. P. Lambropoulos and Mr. P. Maragakis.



\widetext
\narrowtext

\begin{table}
\caption{
Table of the energies of the different ionic thresholds in which the
ionization may leave the Mg atoms, the number of 3.1 eV photons that
need to be absorbed to reach each of these (\(N\mathrm{_{ph}}\)) and
the energy of the ejected electron upon ionization
(\(E\mathrm{_{el}}\))(All numbers refer to field free energies, 
with energy shifts not
taken into account). The various processes are illustrated in 
figure~\ref{fig:mgleveldiagram}.
\label{tbl:mglevels}}
\begin{tabular}{lldld}
No&Transition&Energy (eV)&\(N\mathrm{_{ph}}\)&\(E\mathrm{_{el}}\)(eV)\\
\hline
(1)&Mg(3s$^2$)--Mg$^{+}$(3s)	&7.644	&3	&1.656\\
(2)&Mg(3s$^2$)--Mg$^{+}$(3p)	&12.06	&4	&0.34\\
(3)&Mg$^{+}$(3p)--Mg$^{++}$(2p$^6$)	&10.614	&4	&1.786\\
(4)&Mg$^{+}$(3s)--Mg$^{++}$(2p$^6$)	&15.03	&5	&0.47\\
(5)&Mg(3s$^2$)--Mg$^{++}$(2p$^6)$&22.674	&8	&2.13\\
\end{tabular}
\end{table}

\end{document}